\documentclass[10pt, conference, compsocconf]{IEEEtran}

\usepackage{amsthm}
\usepackage{algorithm}
\usepackage{algorithmic}

\usepackage{cite}
\usepackage{url}
\usepackage{amsmath}
\usepackage{amssymb}
\usepackage{amsthm}
\usepackage{algorithm}
\usepackage{algorithmic}
\usepackage{graphicx}
\usepackage{subfigure}
\usepackage{booktabs}
\usepackage[english]{babel}

\usepackage{ifthen}
\newcommand{\set}[1]{\left\{#1\right\}}
\newcommand{\pr}[1]{\left(#1\right)}
\newcommand{\fpr}[1]{\mathopen{}\left(#1\right)}
\newcommand{\spr}[1]{\left[#1\right]}

\newcommand{\abs}[1]{{\left|#1\right|}}

\newcommand{\enset}[2]{\left\{#1 ,\ldots , #2\right\}}
\newcommand{\enpr}[2]{\pr{#1 ,\ldots , #2}}

\newcommand{\funcdef}[3]{{#1}:{#2} \to {#3}}

\newcommand{\define}{\leftarrow}

\newcommand{\dispfunc}[2]{%
  \ensuremath{%
  \ifthenelse{\equal{\noexpand#2}{}}%
    {{#1}}%
    {{#1}\fpr{#2}}}}

\newcommand{\closure}[1]{cl\fpr{#1}}
\newcommand{\sinks}[1]{sinks\fpr{#1}}
\newcommand{\cover}[1]{\dispfunc{c}{#1}}
\newcommand{\mw}[1]{\dispfunc{m}{#1}}
\newcommand{\nm}[1]{\dispfunc{nm}{#1}}

\newcommand{\lab}[1]{\dispfunc{lab}{#1}}
\newcommand{\sub}[1]{\dispfunc{sub}{#1}}
\newcommand{\parent}[1]{\dispfunc{par}{#1}}

\newcommand{\inc}[1]{\dispfunc{in}{#1}}
\newcommand{\simple}[1]{sm\fpr{#1}}
\newcommand{\co}[1]{co\fpr{#1}}

\newcommand{\mach}[1]{M_{\efam{#1}}}

\newcommand{\efam}[1]{\mathcal{#1}}

\newcommand{\prob}[1]{p\fpr{#1}}
\newcommand{\mean}[2]{\operatorname{E}_{#1}\spr{#2}}

\newtheorem{theorem}{Theorem}
\newtheorem{lemma}[theorem]{Lemma}
\newtheorem{proposition}[theorem]{Proposition}
\newtheorem{corollary}[theorem]{Corollary}

\newtheorem{example}[theorem]{Example}

\begin{document}
\title{Significance of Episodes Based on Minimal Windows}
\author{}
\author{\IEEEauthorblockN{Nikolaj Tatti}
\IEEEauthorblockA{Advanced Database Research and Modelling (ADReM)\\
University of Antwerp, Antwerp, Belgium  \\
nikolaj.tatti@gmail.com}}

\maketitle

\begin{abstract}
Discovering episodes, frequent sets of events from a sequence has been an
active field in pattern mining. Traditionally, a level-wise approach is used to
discover all frequent episodes. While this technique is computationally
feasible it may result in a vast number of patterns, especially when low
thresholds are used.

In this paper we propose a new quality measure for episodes.  We say that an
episode is significant if the average length of its minimal windows deviates
greatly when compared to the expected length according to the independence
model. We can apply this measure as a post-pruning step to test whether the
discovered frequent episodes are truly interesting and consequently to reduce
the number of output.

As a main contribution we introduce a technique that allows us to compute the
distribution of lengths of minimal windows using the independence model.  Such
a computation task is surprisingly complex and in order to solve it we compute
the distribution iteratively starting from simple episodes and progressively
moving towards the more complex ones.  In our experiments we discover candidate
episodes that have a sufficient amount of minimal windows and test each
candidate for significance. The experimental results demonstrate that our
approach finds significant episodes while ignoring uninteresting ones.

\end{abstract}


\section{Introduction}
Discovering episodes, frequent patterns from an event sequence has been a
fruitful and active field in pattern mining since their original introduction
in~\cite{mannila:97:discovery}. Essentially an episode is a set of events that
should occur close to each other with possibly some constraints on the order of
the occurrences.

The most common way of defining a quality measure of an episode is the number
of windows of fixed length in which the episode can be found. Such a measure is
antimonotonic and hence all frequent episodes can be found using
\textsc{APriori} approach given in~\cite{mannila:97:discovery}.  This quality
measure has two significant problems.  First, the results will depend greatly
on the length of the window. If the window is too small, then some interesting
occurrences are ignored. On the other hand, if the window is too large, the
behavior of occurrences in a single window is ignored.

\begin{example}
Consider a serial episode $(a \to b)$ and two sequences '$abababababababab$' and
'$abacbadbaxbagbab$'. If we fix the length of a window to be $6$ (or larger),
then the number of windows covering the episode will be the same for the both
sequences. However, occurrences of the episode in these sequences are different.
\end{example}

The second problem is that this measure has no way of
incorporating any background knowledge. For example, assume that we know that
event $a$ happens relatively seldom, then we are not surprised by the fact if
we observe that an episode containing $a$ also occur seldom.  Alternative
approaches to deal with either the first problem or the second have been
proposed and we discuss them in Section~\ref{sec:related}.

In this paper we propose a new quality measure for the episodes. Our
approach tackles simultaneously both aforementioned problems.  To be more
specific, given an episode $G$, we consider the lengths of minimal windows of
$G$. To include background knowledge we assume that for each symbol we have a
probability of its occurrence in the sequence. We then compute the expected
length of the minimal window based on a model in which the symbols are
independent of each other.  We say that the episode is significant if the
observed minimal windows have abnormal length, that is, the minimal windows are
either too short or too long.

\begin{example}
\label{ex:toy1}
Assume that we have an alphabet of size $3$, $\Sigma = \set{a, b, c}$. Assume
that the probabilities for having a symbol are $\prob{a} = 1/2$, $\prob{b} =
1/4$, and $\prob{c} = 1/4$. Let $G$ be a serial episode $a \to b$.  Then $s$ is
a minimal window for $G$ if and only if it has a form $ac\cdots cb$.  Hence the
probability of a random sequence $s$ of length $k$ to be a minimal window for
$G$ is equal to 
\[
\begin{split}
p(s \text{ is a minimal window  of } G, \abs{s} = k) = \frac{1}{2} \times \frac{1}{4} \times \frac{1}{4^{k - 2}}.
\end{split}
\]
We are interested in a probability of a minimal window having length $k$.
To get this we divide the joint probability by the probability $p(s \text{ is a minimal window  of } G) = 1/6$.
Using this normalization we get that the probability of a minimal window having 
length $k$ is equal to
\[
p(\abs{s}  = k \mid s \text{ is a minimal window of } G) = 3/4 \times 1/4^{k - 2},
\]
for $k \geq 2$, and $0$ otherwise. In this case the distribution is geometric
and the expected length of a minimal window is then $7/3 \approx 2.3$.

On the other hand, assume that we have a sequence $s = accbabacb$. The minimal
windows in $s$ are $s[1, 4]$, $s[5, 6]$, and $s[7, 9]$. Hence, the observed average
length is $(4 + 2 + 3) /3 = 3$.
\end{example}

Computing the probability of the length for a minimal window turns out to be a
surprisingly complex problem. We attack this problem in
Section~\ref{sec:independent} by introducing a certain graph having episodes as
the nodes. Then using this structure we are able to compute the probabilities
inductively, starting from simple episodes and moving towards more complex
ones.

Our recipe for the mining process is as follows: Given the sequence we first
split the sequence in two. The first sequence is used for discovering candidate
episodes, in our case episodes that have a large number of minimal windows (see
Section~\ref{sec:mining} for more details).  Luckily, this condition is
antimonotonic and we can mine these episodes using a standard \textsc{APriori}
method.  We also compute the needed probabilities for the events from the first
sequence. Once the candidate episodes are discovered and the model is computed
we compare the expected length of a minimal window against the average length
of the observed minimal windows from the \emph{second} sequence using a simple
$Z$-test. This step allows us to prune uninteresting episodes, that is, the episodes
that obey the independence model.


The rest of the paper is structured as follows. In
Sections~\ref{sec:prel}--\ref{sec:window} we introduce the preliminary
definitions and notation. In Section~\ref{sec:independent} we lay out our
approach for computing the independence model. We introduce our
method for evaluating the difference between the observed windows and the
independence model in Section~\ref{sec:test}. We discuss mining candidate
episodes in Section~\ref{sec:mining}. Our experiments are given in
Section~\ref{sec:experiments}. We present the related work in
Section~\ref{sec:related} and we conclude our work with discussion in
Section~\ref{sec:conclusions}.

\section{Preliminaries and Notation}
\label{sec:prel}

We begin by presenting preliminary concepts and notations that will be used
throughout the rest of the paper. In this section we will introduce the notions
of sequence and episodes.

A \emph{sequence} $s = \enpr{s_1}{s_L}$ is a string of symbols coming from a
finite \emph{alphabet} $\Sigma$, that is, we have $s_i \in \Sigma$. Such
sequences are generated from random sources, hence we also treat $s$ as a
random variable in our analysis. Given a sequence $s$ and two indices $i$ and
$j$, such that $i \leq j$, we denote by $s[i, j] = \enpr{s_i}{s_j}$ 
a sub-sequence of $s$.

An episode $G$ is represented by an acyclic directed graph with labeled nodes,
that is $G = (V, E, \lab{})$, where $V = \enpr{v_1}{v_K}$ is the set of nodes, $E$
is the set of directed edges, and $\lab{}$ is the function $\funcdef{\lab{}}{V}{\Sigma}$,
mapping each node $v_i$ to its label.

Given a sequence $s$ and an episode $G$ we say that $s$ \emph{covers}  the
episode if there is an \emph{injective} map $f$ mapping each node $v_i$ to a valid
index such that the node and the corresponding sequence element have the same
label, $s_{f(v_i)} = \lab{v_i}$, and that if there is an edge $(v_i, v_j) \in E$, then
we must have $f(v_i) < f(v_j)$. In other words, the parents of the node $v_i$
must occur in $s$ before $v_i$. We define a binary function $\cover{s; G}$
such that $\cover{s; G} = 1$ if and only if $s$ covers $G$.
Traditional episode mining is based on searching episodes that are covered by
sufficiently many sub-windows of certain fixed size.

An elementary theorem says that in directed acyclic graph there exists a sink,
a node with no outgoing edges. We denote the set of sinks by $\sinks{G}$.
Given an episode $G$ and a node $v$, we define $G - v$ to be the sub-episode
obtained from $G$ by removing $v$ (and the incident edges).

Given a collection of episodes $\efam{G}$ we say that the collection is
downward closed, if for a given $G \in \efam{G}$, each subgraph $H$ of $G$ is
included in $\efam{G}$. Note that the empty episode is included in $\efam{G}$.
Throughout the whole paper we will be working with downward closed
collections of episodes $\efam{G}$.

\section{Minimal Windows of Episodes}
\label{sec:window}
Episode mining is based on finding episodes that occur often enough in sliding
window. We approach the problem from a different angle. Given a sequence and a
candidate episode we first discover the set of all minimal windows in which the
given episode occurs.  Once we have obtained this set we will study the length
of these windows.  If their distribution is abnormal, either the lengths are
too short, or too long, we consider that we have discovered an important
episode.

In order to make the preceding discussion more formal, let $G$ be an episode,
and let $s$ be a sequence. We say that $s$ is a \emph{minimal window} for $G$
if $G$ is covered by $s$ but not by any proper sub-window of $s$.  We define
a function $\mw{s; G}$ returning a binary value. The function $\mw{s; G} = 1$
if and only if $s$ is a minimal window starting for $G$.

Let $s$ be a random sequence of length $L$, and write $t = s[1, L - 1]$ and $u
= s[2, L]$. Note that we can write the probability of $s$ being a minimal window as
\begin{equation}
\label{eq:minwin}
\begin{split}
\prob{\mw{s; G}} & = \prob{\cover{s; G}} - \prob{\cover{t; G} \lor \cover{u; G}} \\
  &= \prob{\cover{s; G}} - \prob{\cover{t; G}}  - \prob{\cover{u; G}} \\
 & \quad + \prob{\cover{t; G}, \cover{u; G}}.
\end{split}
\end{equation}

Our main focus is the distribution of lengths of minimal windows, that is, we
are interested in
\[
	p_G(k) = p\fpr{\abs{s} = k \mid \mw{s; G} = 1},
\]
where $s$ is a random sequence. Note that this distribution is
defined for all $k$. In practice, we compute $\prob{\mw{s; G}}$ for $k = 1,
\ldots, K$, where $K$ is some suitable predetermined constant. Once these
values are computed we normalize them so that $p_G(k)$ becomes a proper
distribution. This is equivalent to saying that we are not interested in
minimal windows whose length exceeds $K$. From now on $K$ will always denote
the maximal length of a minimal window.

We should point out that even though we limit ourselves to windows of maximal
size $K$ this limitation is not as severe as using windows of fixed size $K$.
While we ignore information of longer windows we still are able to detect any
deviation occurring with the length of minimal windows shorter or equal than
$K$. On the other hand, in the fixed window approach the information of the
length of minimal window is discarded as long as it is short enough.

\section{Computing the Minimal Windows from Independence Model}
\label{sec:independent}
We devote this section for computing the distribution of lengths of minimal
windows. That is we are given probabilities $p(a)$ for each symbol in an
alphabet and a set of episodes $\efam{G}$ and for each $G \in \efam{G}$ we wish
to compute $p_G(k)$, the probability of a minimal window of $G$ being of size
$k$, according to the independence model.

Our approach for calculating minimal windows is based on Eq.~\ref{eq:minwin}.
According to this equation we need to solve the probabilities $\prob{\cover{s;
G}}$ and $\prob{\cover{s[1, L - 1]; G}, \cover{s[2, L]; G}}$. We will achieve
this by building certain finite state machines where the states will correspond
to the episodes. 

\subsection{Episode Set as Finite State Machine}
It will be fruitful to represent the given set of episodes $\efam{G}$ as a
certain finite state machine. To be more precise, we define a finite state
machine as a DAG $\mach{\efam{G}} = \pr{V(\mach{\efam{G}}),
E(\mach{\efam{G}})}$. The states $V(\mach{\efam{G}})$ are exactly the episodes
$\efam{G}$.  Let $X, Y \in \efam{G}$ be two episodes and let $v$ and $w$ be the
corresponding states.  An edge $e = (v, w) \in E(\mach{\efam{G}})$ with a label
$a = \lab{e}$ exists if and only if $X = Y - n$, where $n$ is a sink node of
$Y$ labeled as $a$.  In other words, there should be an edge between an episode
and an episode obtained by removing a sink node with a label $a$.

\begin{example}
\label{ex:machtoy1}
We consider a downward set of closed episodes $\efam{G} = \set{(a, b \to a), (b \to
a), (a, b), (a), (b), (a, a), \emptyset}$. The machine $\mach{\efam{G}}$ is
given in Figure~\ref{fig:toy1}.
\begin{figure}
\center
\includegraphics[width=5.5cm]{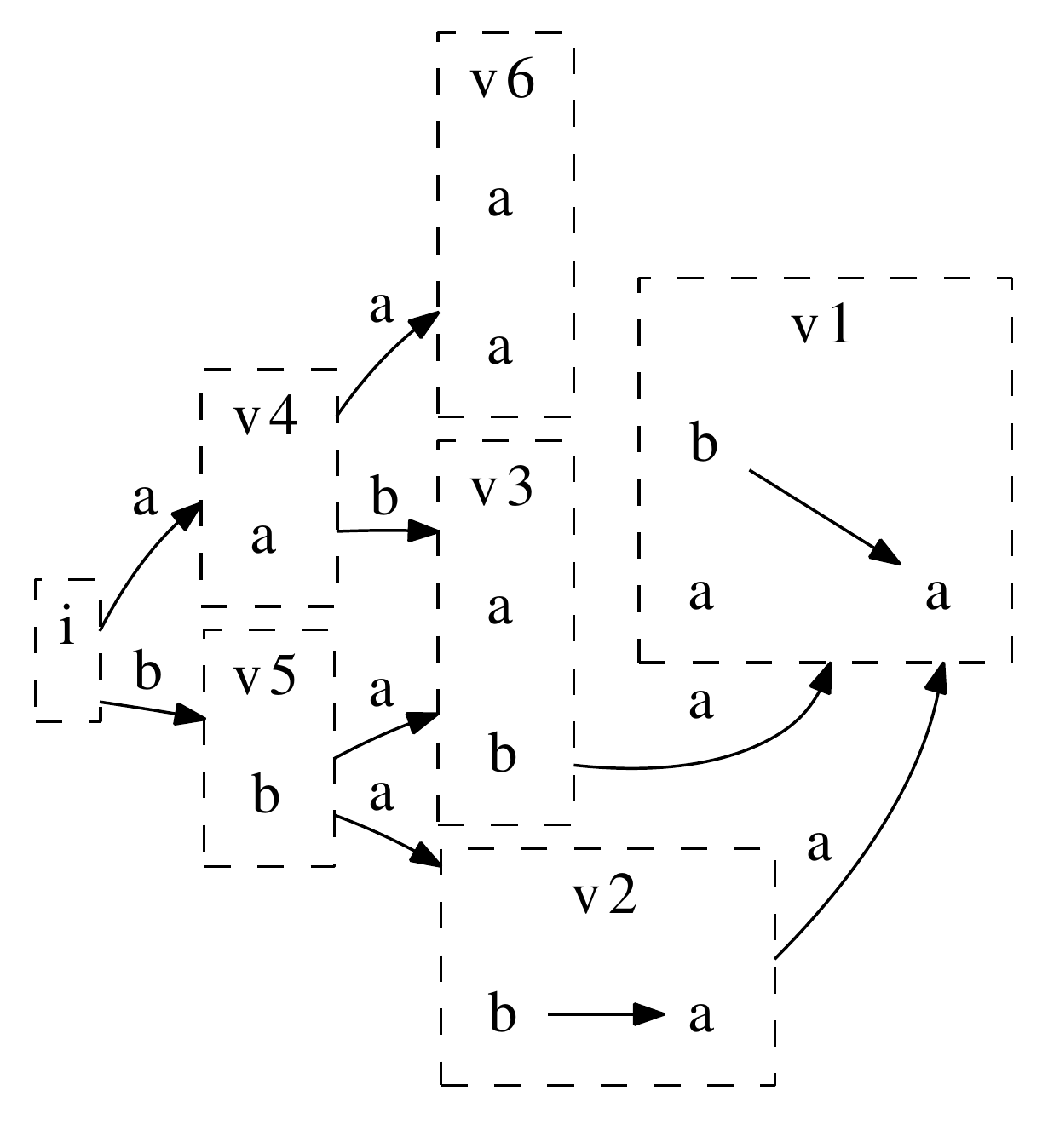}
\caption{The machine $\mach{\efam{G}}$ for the episodes $\efam{G} = \set{(a, b
\to a), (b \to a), (a, b), (a), (b), (a, a), \emptyset}$. Each edge represents
a removed sink between the episode and the parent episode.}
\label{fig:toy1}
\end{figure}
\end{example}

Given a state $v$ in $\mach{\efam{G}}$ we say that $s$ \emph{covers} $v$ if
there is a sequence $t = \enset{s_{i_1}}{s_{i_N}}$ such that $v$ can be reached
when $t$ is given as an input. In that case we set $\cover{s ; v} = 1$, and $0$
otherwise. Similarly we define $\mw{s ; v}$, when $s$ covers $v$ but $s[2, L]$
and $s[1, L - 1]$ does not cover $v$.

Comparing this to the definition of coverage for the episode we see the
immediate result.
\begin{proposition}
\label{prop:cover}
The sequence $s$ covers an episode $G \in \efam{G}$ if and only if $s$ covers
the corresponding $v$ in $\mach{\efam{G}}$. Consequently, a minimal window of
$G$ is equivalent to the minimal window of $v$.
\end{proposition}

Let $M$ be a finite state machine and let $v$ be a state in $M$ We say that $v$
is \emph{monotonic} if a sequence $s$ covering $v$ also covers any parent state of
$v$. If every state in $M$ is monotonic we say that $M$ is \emph{monotonic}.

A direct corollary of Proposition~\ref{prop:cover} states that
$\mach{\efam{G}}$ is monotonic.

\begin{lemma}
\label{lem:monotonic1}
The machine $\mach{\efam{G}}$ induced from $\efam{G}$ is monotonic.
\end{lemma}

\begin{IEEEproof}
Let $v$ a state in $\mach{\efam{G}}$ and let $w$ be its parent state.  Let $X$
be the episode represented by the state $v$ and let $Y$ be the episode
represented by the state $w$. If $v$ covers $s$, then it must cover $X$. Since
$Y$ is an episode obtained from $X$ by removing one sink, $s$ also covers $Y$
and thus cover $w$.
\end{IEEEproof}

\subsection{Computing Coverage for States of Simple Machines}
In this section we will demonstrate how to compute the probabilities
$\prob{\cover{s ; v}}$ for the state $v$ in $M$. We will make some simplifying
assumption concerning the structure of $M$, and then in the next section we
demonstrate how this limitations can be removed.

We say that a machine $M$ is \emph{simple} if incoming edges for each state $v$
in $M$ have unique labels. Generally, the episode machine $\mach{\efam{G}}$ is not simple.
However, if the episodes $\efam{G}$ have only nodes with unique labels, then
$\mach{\efam{G}}$ will be simple.

Our approach is to compute the coverage of a state $v$ based on the coverage of its
parent state.

\begin{proposition}
Let $M$ be simple and monotonic. Let $v$ be a state in $M$, and let $s$ be a
random sequence of length $L$ with independent symbols. Let $t = s[1, L - 1]$ be the sub-sequence of $s$
without the last element. Define probability $d$ as $\prob{\cover{s; v}} = d + \prob{\cover{t; v}}$.
Then
\[
	d = \sum_{e = (w, v) \in E(M)} p\fpr{\lab{e}}\fpr{\prob{\cover{t; w}} - \prob{\cover{t; v}}}.
\]
\end{proposition}

\begin{IEEEproof}
By definition, we have $d = p\fpr{\cover{s; v}} - p\fpr{\cover{t; v}}$, that
is, $d$ is the probability of sequence $s$ covering $v$ but $t = s[1, L - 1]$
not covering it. Assume that $s$ is such sequence. This implies that there is
an edge $e =(w, v)$ with $\lab{e} = s_L$. Fix $s_L = \lab{e}$. Since $M$ is simple the only
path to reach $v$ must use the unique $e$. 
We also must have that $t$ covers $w$ but not $v$. Note that
this probability can be written as
\[
p(\cover{t; w}) - p(\cover{t; w}, \cover{t; v}) = p(\cover{t; w}) - p(\cover{t; v}),
\]
where the equality follows from since $M$ is monotonic.

The probability of $s_L$ being $\lab{e}$ is $\prob{\lab{e}}$. The result follows by
combining these probabilities.
\end{IEEEproof}

Let us abuse the notation and write $\prob{\cover{L; v}}$ to mean the
probability $\prob{\cover{s; v}}$ where $s$ is a random sequence of length $L$.
The proposition gives us means to compute $\prob{\cover{L; v}}$ in an iterative
fashion from $\prob{\cover{L - 1; v}}$ and from the coverage of parent state.
The algorithm for computing the coverage is given in Algorithm~\ref{alg:cover}.

\begin{algorithm}
\begin{algorithmic}[1]
\FOR{$e = (w, v) \in E(M)$}
	\IF {coverage for $w$ is not computed}
		\STATE $\textsc{CoverState}(w)$. 
	\ENDIF
\ENDFOR

\FOR{$k = 1, \ldots K$.}
	\STATE $d \define 0$.
	\FOR{$e = (w, v) \in E(M)$}
		\STATE $x \define \prob{\cover{k - 1; w}} - \prob{\cover{k - 1; v}}$.
		\STATE $d \define d + p\fpr{\lab{e}}x$.
	\ENDFOR
	\STATE $\prob{\cover{k; v}} \define d + \prob{\cover{k - 1; v}}$.
\ENDFOR

\end{algorithmic}
\caption{Recursive sub-procedure \textsc{CoverState} for computing the coverage of state $v$.}
\end{algorithm}

\begin{algorithm}
\begin{algorithmic}[1]
\STATE $s \define$ the source state of $M$.
\STATE $\prob{\cover{k; s}} \define 1$, for $k = 0, \ldots, K$.
\FOR{$v$ sink state in $M$}
	\STATE $\textsc{CoverState}(v)$. 
\ENDFOR

\end{algorithmic}
\caption{Algorithm \textsc{Cover} for computing the coverage of state for a
simple and monotonic machine $M$.}
\label{alg:cover}
\end{algorithm}

To analyze the computational complexity, we first note that computing the
coverage of state $v$ requires $O(KL)$ steps where $L$ is the number of
incoming edges of $v$. Thus computing the coverage of the complete graph will
require $O(\abs{E(M)}K)$ steps. However, in practice the process is more
complex.  The computations are not numerically stable due to rounding errors in
floating-point numbers. To solve this problem we have to resort to exact
rational numbers. Using such numbers implies that simple computations are no
longer unit operations making the computation times longer.

\subsection{Transforming non-simple Machines}

In order to use Algorithm~\ref{alg:cover} our state machine needs to be simple
and monotonic. The machine $\mach{\efam{G}}$ is monotonic but not simple.
Luckily, we can define a new simple and monotonic machine from which we can
compute the coverage.  Informally, if we reverse the direction of the edges in
$\mach{\efam{G}}$, then making the machine simple is equal to making the
reversed non-deterministic machine deterministic.

In order to make this formal, let us first give some definitions. Let $V$
be a subset of states in $M$. Let $a$ be a label.  We also define
\[
\begin{split}
\sub{V; a} = \{w  \mid & e = (w, v) \in E(M), \lab{e} = a, v \in V \}
\end{split}
\]
to be the set of parents of each $v \in V$ connected with an edge having a label $a$.
We define 
\[
\parent{V; a} = \min \fpr{\sub{V; a} \cup V},
\]
where $\min\fpr{X}$ results in minimal states of $X$ with respect to the
parenthood in $M$. This guarantees that $\parent{V; a}$ contains no state $v, w$
such that $v$ is an ancestor of $w$. We also need to define
\[
\inc{V} = \set{\lab{e} \mid e = (w, v) \in E(M), v \in V}
\]
to be the set of labels of all incoming edges.  We define the closure of $V$
inductively to be the collection of sets of states
\[
\closure{V} = \set{V} \cup \bigcup_{a \in \inc{V}} \closure{\parent{V; a}}
\]
and $\closure{V} = \set{V}$ if $\inc{V}$ is empty.

We are now ready to transform $M$ into a simple machine, which we denote by $\simple{M}$.
The states of the machine $\simple{M}$ are
\[
V(\simple{M}) = \bigcup_{v \in V(M)} \closure{\set{v}}.
\]
Let $V$ be a state in $\simple{M}$ and let $a \in \inc{V}$. Let $W = \parent{V; a}$. Note that
this state exists in $\simple{M}$.  We define an edge labeled as $a$ from
$W$ to $V$. Let $i$ be the initial state in $M$. For $\mach{\efam{G}}$ it is
the state corresponding to the empty episode. Then $\set{i}$ is the initial
state for $\simple{M}$. From now on $\set{i}$ will always denote the initial
state of $\simple{M}$.

\begin{example}
\label{ex:machtoy2}
We continue Example~\ref{ex:machtoy1}. Note that $\mach{\efam{G}}$ is not simple
since $v_1$ has two incoming edges with a label $a$.  The transformed machine
$\simple{\mach{\efam{G}}}$ is given in Figure~\ref{fig:toy2}. Note that, in
addition to the states already existing in $\mach{\efam{G}}$ is has now two
extra states, namely $\set{v_2, v_3}$ and $\set{v_2, v_4}$. Also note that
$\simple{\mach{\efam{G}}}$ is simple.
\begin{figure}
\center
\includegraphics[width=5.5cm]{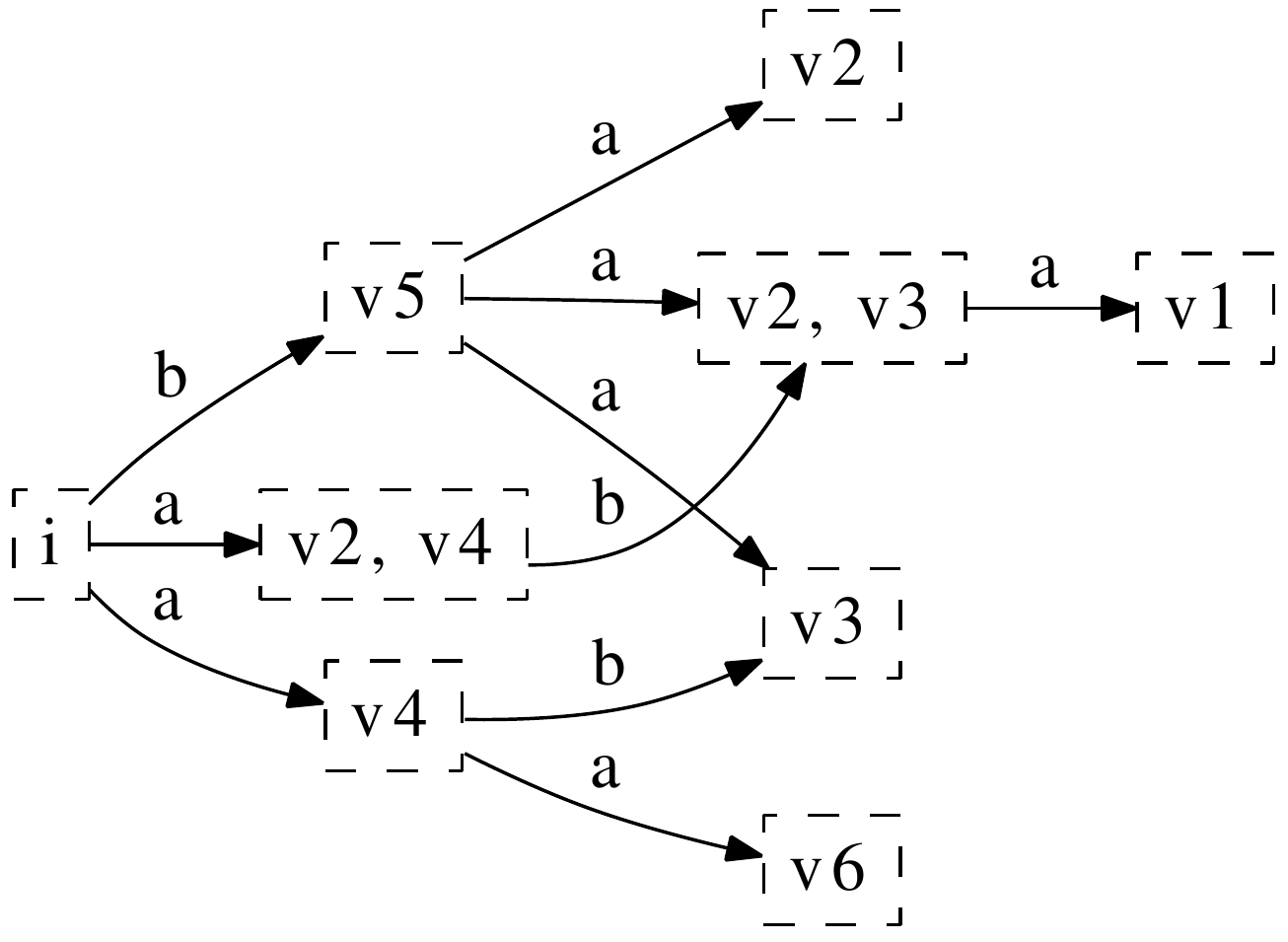}
\caption{The transformed machine $\simple{\mach{\efam{G}}}$ of the machine
$\mach{\efam{G}}$ given in Figure~\ref{fig:toy1}.}
\label{fig:toy2}
\end{figure}
\end{example}

It is obvious that $\simple{M}$ is simple. The following proposition reveals the
expected relationship between $M$ and $\simple{M}$.

\begin{proposition}
Let $M$ be a monotonic machine.  Let $V = \enset{v_1}{v_N}$ be the state in
$\simple{M}$. Then a sequence $s$ covers $V$ if and only if $s$ covers at least
one $v_i$.
\end{proposition}

\begin{IEEEproof}
We will prove this by induction. Assume that the proposition holds for all parent
states of $V$.

Assume that $s$ covers $V$.  Let $t$ be a sub-sequence of $s$ that leads
$\simple{M}$ from the source state $\set{i}$ to $V$.  Let $s_e$ be the last
symbol of $s$ occurring in $t$.  There is a parent state $W = \enset{w_1}{w_L}$
which $s[1, e - 1]$ covers.  By the induction assumption at least one $w_k$ is
covered by $s[1, e - 1]$. If there is $v_j = w_k$, then $v_j$ is covered by
$s$, otherwise there is $v_j$ that has $w_k$ as a parent state.  The edge
connecting $v_j$ and $w_k$ has a label $s_e$. Hence $s$ covers $v_j$ also.

To prove the other direction assume that $s$ covers $v_j$. Let $t$ be a
sub-sequence that leads $M$ from the source state to $v_j$. Let $s_e$ be the
last symbol occurring in $t$. Let $w$ be the parent state of $v_j$ connected by
an edge with a label $s_e$. Since $s_e \in \inc{V}$, we must have $W$ as a
parent state of $V$ such that either $w \in W$ or an ancestor state $u$ of $w$
is in $W$.  In the latter case, since  $M$ is monotonic, $s$ covers $u$.  In
either case, by the induction assumption $s[1, e - 1]$ covers $W$. Hence $s$
covers $V$.
\end{IEEEproof}

\begin{corollary}
Sequence $s$ covers $v$ in $M$ if and only if $s$ covers $\set{v}$ in
$\simple{M}$.
\end{corollary}

\begin{corollary}
Let $M$ be a monotonic machine, then $\simple{M}$ is also monotonic.
\end{corollary}

\begin{IEEEproof}
Let $V$ be a state in $\simple{M}$ and let $s$ cover $V$. Let $W$ be a  parent
state of $V$. Let $v \in V$ such that $s$ covers $v$. If $v \in W$, then $s$
covers $W$. Otherwise there is an ancestor state $w \in W$ of $v$. Since $M$ is
monotonic, $s$ covers $w$ and thus $W$.
\end{IEEEproof}

The corollaries give us means to compute the coverage of states in
$\mach{\efam{G}}$ by solving the coverage of the states $\simple{\mach{\efam{G}}}$
using Algorithm~\ref{alg:cover}.

\subsection{Computing Co-coverage}
Our last challenge is to compute the term $\prob{\cover{t; G}, \cover{u; G}}$
in Eq.~\ref{eq:minwin}. In order to do that we design a special finite state
machine, denoted by $\co{M}$, in which the coverage of certain states will
correspond to the last term in Eq.~\ref{eq:minwin}.  The construction of this
machine is based on the previous machines $M = \mach{\efam{G}}$ and
$\simple{M}$. To avoid confusion we use $v$ and $w$ for the states in $M$,
$V$ and $W$ for the states in $\simple{M}$, and greek letters $\alpha, \beta,
\ldots$ for the states in $\co{M}$. We proceed by constructing the machine
first and then prove that it gives us the desired probabilities.

There are three different kinds of states in $\co{M}$. The first group consists
of one state, namely $\eta = \set{i}$, where $\set{i}$ is the initial state of
$\simple{M}$. This state will be the initial state of $\co{M}$.

The second group consists of certain pairs of states from $\simple{M}$. Let $V$
and $W$ be states in $\simple{M}$ and write $\alpha = \pr{V, W}$. The machine
will be constructed in such manner that $s$ will cover $\alpha$ in $\co{M}$ if
and only if $s$ covers $V$ and $s[2:L]$ covers $W$ in $\simple{M}$. In order to
achieve this we first define a closure
\[
\closure{\alpha} = \set{\alpha} \cup \bigcup_{a \in \inc{V} \cup \inc{W} } \closure{\pr{\parent{V; a}, \parent{W; a}}}.
\]
Let $v$ be a state in $M$ that is not the source state.  For each label $a \in
\inc{\set{v}}$ we add the states from $\closure{\pr{\set{v}, \parent{\set{v} ;
a}}}$ into $\co{M}$. In addition, we add the states from $\closure{\pr{\set{v}, \set{v}}}$.
We add an edge with a label $a$ to $\alpha = (V_1, V_2)$ from $\beta = (W_1,
W_2)$ if $W_k = \parent{V_k; a}$ for $k = 1, 2$ with one exception: if
$\parent{V_1 ; a} = V_2 = \set{i}$, then instead if connecting $\alpha$ to
$(\set{i}, \set{i})$ we connect $\alpha$ to $\eta = \set{i}$.
We also connect the state $(\set{i}, \set{i})$ to $\eta$ with an edge accepting
any symbol from the alphabet. See Example~\ref{ex:machtoy3} for illustration.

We will now define our last group of states. For any state $v$ that is not a
source in $M$ we add a state $\alpha = v$. For each $a \in \inc{\set{v}}$, we
add an edge with a label $a$ from $\pr{\set{v}, \parent{\set{v} ; a}}$ to $\alpha$.
We also add an edge from $\pr{\set{v}, \set{v}}$ to $\alpha$ accepting any symbol
outside $\inc{\set{v}}$.

\begin{example}
We continue the toy example given in Example~\ref{ex:machtoy2}. A part of
the machine $\co{\mach{\efam{G}}}$ is given in Figure~\ref{fig:toy3}. Namely
we show the machine solving the co-coverage for the episodes $v_2 = (b \to a)$
and $v_5 = (b)$.
\label{ex:machtoy3}
\begin{figure}
\includegraphics[width=\columnwidth]{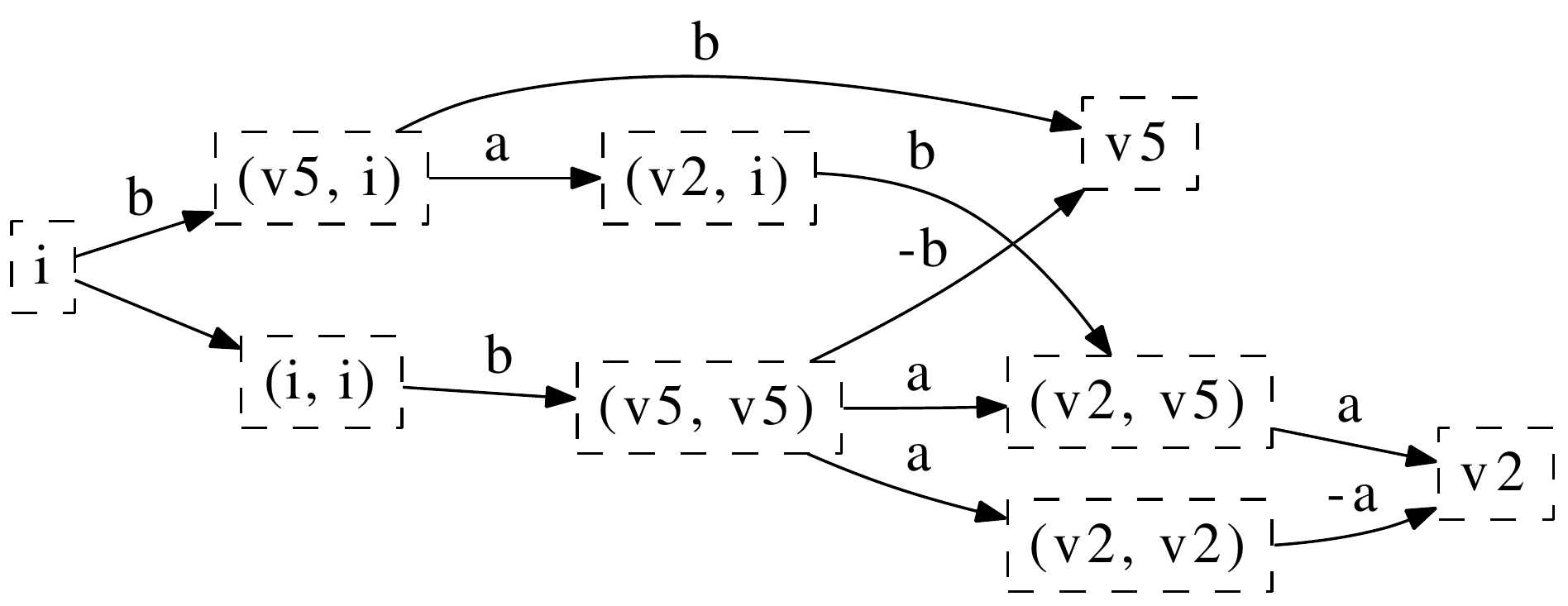}
\caption{A part of $\co{\mach{\efam{G}}}$. Here we have only included states
corresponding to the episodes $(b)$ and $(b \to a)$. The edge $-a$ means that
the edge accepts any symbol but $a$.}
\label{fig:toy3}
\end{figure}
\end{example}

Now that we have defined our machine we are ready to prove that the coverage of
the states of the last group actually corresponds to the last term in
Eq~\ref{eq:minwin}.

First, we need to point out a certain property of $\simple{M}$.

\begin{lemma}
\label{lem:parent}
Let $M$ be a monotonic machine. Let $s$ be a sequence of length $L$ covering a
state $V$ in $\simple{M}$. Then $s[1, L - 1]$ covers $\parent{V; s_L}$.
\end{lemma}

\begin{IEEEproof}
If $s_L \notin \inc{V}$, then a sub-sequence $t$ leading to $V$ does not
contain $s_L$.  Hence $s[1, L - 1]$ covers $V = \parent{V; s_L}$. Assume that
$s_L \in \inc{V}$. Let $W = \parent{V; s_L}$. If $s[1, L - 1]$ covers $V$, then
by the monotonicity of $\simple{M}$, $s$ also covers $W$. If $s[1, L - 1]$ does
not cover $V$, then $t$ must have $s_L$ as a last symbol and since $\simple{M}$
is simple, $t$ must go through $W$ Hence, $s[1, L - 1]$ covers $W$.
\end{IEEEproof}

Our second step is to describe the coverage of the intermediate states in
$\co{M}$.

\begin{proposition}
\label{prop:costate}
Let $s$ be a sequence of length $L$.  Let $\alpha = (V_1, V_2)$ be a state in
$\co{M}$, then $s$ covers $\alpha$ if and only if $s$ covers $V_1$ and $s[2,
L]$ covers $V_2$.
\end{proposition}

\begin{IEEEproof}
We will prove the result by induction.  To prove the first step assume that
$\parent{V_1; a} = \parent{V_2; a} = \set{i}$. If $V_2 = \set{i}$, then $V_1$
is connected to the state $\eta = \set{i}$ (in $\simple{M}$) by an edge with a
label $a$. Hence, $s$ covering $\alpha$ is equivalent to $s$ covering $V_1$.
The result follows since the state $V_2 = \set{i}$ is automatically covered.
Assume now that $V_2 \neq \set{i}$.  In this case $s[2, L]$ covering $V_2$
implies that $s[2,L]$ (and hence also $s$) covers $V_1$.  Since $\alpha$ is
connected to $(\set{i}, \set{i})$, $s$ covering $\alpha$ is equivalent that
$s[2, L]$ has a symbol $a$. But this is equivalent for $s[2, L]$ covering
$V_2$.

Assume now that the result holds for all parent states of $\alpha$. Assume that
$s$ covers $\alpha$. There must be a symbol $s_e$ and a parent state $\beta =
(W_1, W_2)$ linked to $\alpha$ by an edge with a label $s_e$ such that $s[1, e
- 1]$ covers $\beta$.  By the assumption $s[1, e - 1]$ covers $W_1$ and $s[2, e
  - 1]$ covers $W_2$.  Thus, $s[1, e]$ covers $V_1$ and $s[2, e]$ covers $V_2$.

Assume now that $s$ covers $V_1$ and $s[2, L]$ covers $V_2$. Let $s_e$ be the
last element in $s$ such that $s_e \in \inc{V_1}$ or $s_e \in \inc{V_2}$. By
Lemma~\ref{lem:parent} we must have that $s[1, e - 1]$ covers $\parent{V_1;
s_e} = W_1$ and $s[2, e - 1]$ covers $\parent{V_2; s_e} = W_2$.  Hence by the
induction assumption $s[1, e - 1]$ covers $\beta = (W_1, W_2)$ and
consequently, since $\beta$ is a parent state of $\alpha$, $s[1, e]$ covers
$\alpha$.
\end{IEEEproof}

Now we are ready to prove that the coverage of states in the last group is exactly
what we wish to have.

\begin{proposition}
Let $\alpha$ be a state in $\co{M}$ corresponding to a state $v$ in $M$.  Then
a sequence of length $L$ covers $\alpha$ if and only if $s[1, L - 1]$ and $s[2,
L]$ cover $v$.
\end{proposition}

\begin{IEEEproof}
Assume that $s$ covers $\alpha$. Let $t$ be a sub-sequence of $s$ that leads to
$\alpha$ from the source state. Let $s_e$ be a last symbol of $t$. If $s_e \in
\inc{\set{v}}$ then $t$ travels through $\beta = (\set{v}, \parent{\set{v};
s_e})$.  By Proposition~\ref{prop:costate}, $s[1, e - 1]$ covers $\set{v}$ and
$s[2, e - 1]$ covers $\parent{\set{v}; s_e}$ and hence $s[2, e]$ covers $v$. If
$s_e \notin \inc{\set{v}}$, then $t$ travels through $(\set{v}, \set{v})$ and
result follows again from Proposition~\ref{prop:costate}.

To prove the other direction assume now that $s[1, L - 1]$ and $s[2, L]$ cover
$\set{v}$. Assume that $s_L$ is not in $\inc{\set{v}}$. This implies that $s[2, L -
1]$ covers $\set{v}$. Thus, by Proposition~\ref{prop:costate}, $s[1, L - 1]$ covers
$(\set{v}, \set{v})$ and consequently $s$ covers $\alpha$. On the other hand,
if $s_L \in \inc{\set{v}}$, then Lemma~\ref{lem:parent} implies that $s[2, L -
1]$ covers $\parent{\set{v}; s_L}$ and hence must cover, by
Proposition~\ref{prop:costate}, $(\set{v}, \parent{\set{v}; s_L})$.
Consequently $s$ covers $\alpha$.
\end{IEEEproof}

It is easy to see that $\co{M}$ is simple and monotonic, if $M$ is monotonic.
Hence, we can compute the coverage of $\co{\mach{\efam{G}}}$ using
Algorithm~\ref{alg:cover}.

We can now compute the probability of $s$ being a minimal window using
Eq.~\ref{eq:minwin}. First we solve the coverage using
$\simple{\mach{\efam{G}}}$.  Secondly, we compute the co-coverage using
$\co{\mach{\efam{G}}}$. Once these are computed we can use Eq.~\ref{eq:minwin}.

Let us finish by discussing the relative sizes of the machines. It can be shown
that the number of states in $\simple{\mach{\efam{G}}}$ can be substantially
larger than the number of states $\mach{\efam{G}}$. Consequently, in the worst
case our method is not polynomial. Such an explosion, however, requires a
specific episode with many nodes having the same label. Such episodes are
unlikely to be candidates if we are dealing with sequences that have a large
alphabet distributed more or less evenly.  Moreover, we demonstrate later that
in our experiments the size of $\simple{\mach{\efam{G}}}$ is about the same as
the size of $\mach{\efam{G}}$.

\section{Testing Minimal Windows}
\label{sec:test}
In this section we will describe how we test whether the discovered minimal
windows obey the independence model.

We say that the episode is significant if the average length of the minimal
windows is abnormally small or large. In order to measure the abnormality we
will use a $Z$-test.  In order to perform this test we need to show that the
average length is asymptotically normal and compute the mean and the variance
according to the independence model. This is not trivial since the minimal
windows correlate within a single sequence $s$.  For example, assume that we
are looking for the parallel episode $(a, a, a)$, and assume that we have found
a minimal window of size $3$, that is, the window is $aaa$. Then the next
minimal window will have a higher probability of being short, since we already
have two $a$s.  Thus the occurrences of minimal windows in $s$ are not
independent even if $s$ obeys the independence model.

Assume that we are given a long random sequence $s$ of length $N$ and 
write $X_i$ to be a boolean random variable such that $X_i = 1$ if there is a
minimal window starting at $i$th symbol. Also let $Y_i$ be the length of that
minimal window and $0$ if there is no window. Note that the estimator of the
average length is
\[
M = \sum_i Y_i / \sum_i X_i.
\]
Let us first show that $M$ is normally distributed. To see that note that
$(Y_i, X_i)$ and $(Y_{i + K}, X_{i + K})$, where $K$ is the maximum length of a
window, are independent. Hence, $(X_i, Y_i)$ is a \emph{strongly mixing} sequence
which allows us to use a Central Limit Theorem for dependent variables (given
in~\cite{billingsey:95:probability}, for example) so that $(\sqrt{N}X_i,
\sqrt{N}Y_i)$ is asymptotically normal as $N$ approaches infinity. Let us
denote by $C$ the covariance matrix of this limit distribution.
Also write $p = \mean{}{X_1}$ and $q = \mean{}{Y_1}$.
Using the same theorem we know that the components of $C$ are
\begin{eqnarray*}
	C_{11} & = &  \mean{}{(Y_1 - q)^2} + 2\sum_{i = 2}^K \mean{}{(Y_1 - q)(Y_i - q)} \\
	C_{22} & = &  \mean{}{(X_1 - p)^2} + 2\sum_{i = 2}^K \mean{}{(X_1 - p)(X_i - p)} \\
	C_{12} & = &  \mean{}{(X_1 - p)(Y_1 - q)} + \\
	       &  & \sum_{i = 2}^K \mean{}{(X_1 - p)(Y_i - q) + (Y_1 - q)(X_i - p)}. \\
\end{eqnarray*}

Let us define $m = q/p$ which is the average length of the minimal window.
Since $f(y, x) = y/x$ is continuous and differentiable function at $(q, p)$, we
know from Theorem~3.1 in~\cite{vaart:98:asymptotic} that $\sqrt{N}f(\sum Y_i,
\sum_i X_i)$ is asymptotically normal with mean $m$ and variance
$\sigma^2$, where
\[
	\sigma^2 = \nabla f^T C \nabla f = \frac{1}{p^2}\pr{C_{22} - 2mC_{12} + m^2C_{11}},
\]
where $\nabla f = (1/p, -m/p)$ is the gradient of $f$ at $(q, p)$.

Thus in order to perform a  statistical test for an episode $G$ given a sequence
$s$, let $W$ be the sum of lengths of the discovered minimal windows. We consider
the following statistic
\begin{equation}
\label{eq:ztest}
	Z = \frac{W - Nm}{\sqrt{N}\sigma}.
\end{equation}
Based on the above discussion if $s$ truly comes from the independence model, then
$Z$ is asymptotically distributed as a standard normal distribution $N(0, 1)$.

Our remaining task is to compute $m$ and $\sigma$. Note that since we know the
probability of $s$ being a minimal window, we can compute $p$, $q$, $m$, and
the first terms of $C_{11}$, $C_{12}$, and $C_{22}$. However the last terms of
$C$ cannot be computed easily. We resolve this issue by simulating a sequence
of independence model and estimating these terms from that sequence. 

\section{Mining Candidate Episodes with non-overlapping Minimal Windows}
\label{sec:mining}
So far we have assumed that we already know what episodes we wish to test.
In this section we will focus on mining candidate episodes.

In order to have a reliable $Z$-statistic (see Eq.~\ref{eq:ztest}), we need to
have a decent number of minimal windows. Hence, a good criterion for a candidate
episode is that the number of minimal windows exceeds some given threshold. This
is the criterion used in \textsc{Minepi} (see~\cite{mannila:97:discovery}).
However, this condition is not antimonotonic as demonstrated in the next toy
example.

\begin{example}
Consider the sequence '$aba$'. There are $2$ minimal windows for the parallel
episode $(a, b)$, yet there is only one minimal window for the episode $(b)$.
\end{example}

We remedy this problem by making a stronger requirement. We search all the
episodes whose number of  non-overlapping windows exceed some given threshold.
It turns out, that this condition is antimonotonic and we can search the
episodes in a level-wise fashion.

Since there are several ways of selecting non-overlapping subcollection of
minimal windows, we will give a more precise definition. Let $W$ be a sequence
$W$ of minimal windows of an episode $G$ in a sequence $s$. Assume that the
minimal windows in $W$ are ordered by their occurrences in $s$. We select the
first window and remove any window that overlaps with the selected window.  We
repeat this until the $W$ has no more windows. We define $\nm{G; s}$ to be the
minimal windows discovered in such fashion. We first show that this approach
produces the maximal number of samples.

\begin{proposition}
\label{prop:monotonic}
Let $V$ be a collection of non-overlapping minimal windows of an episode $G$ in
a sequence $s$. Then $\abs{V} \leq \abs{\nm{G; s}}$.
\end{proposition}

\begin{IEEEproof}
Let $W$ be the collection of possibly overlapping minimal windows of $G$ in
$s$.  We will prove that among any sub-collection of $W$ of non-overlapping
windows, the collection $\nm{G; s}$ has the maximal size.  We will prove this
by induction over the size of $W$.

Let $w \in W$ be the first window in $W$. Let $X$ be the set of windows that
overlap with $w$ (note that $w \in X$). By the definition, we have $w \in
\nm{G; s}$, and the next window will be the first window outside $X$.

If $V \cap X = \emptyset$, then $V \subseteq W - X$, and the result follows
from the induction assumption. Assume that $V \cap X \neq \emptyset$.  Any two
windows $x,y \in X$ must overlap, hence $V$ can contain exactly one member of
$X$, say $x$. This means that $V - \set{x} \subseteq W - X$, and the result
follows from the induction assumption. 
\end{IEEEproof}

\begin{corollary}
The quantity $\abs{\nm{G; s}}$ is antimonotonic.
\end{corollary}
\begin{IEEEproof}
Let $H$ be a sub-episode of $G$.  Then any minimal window in $\nm{G; s}$ also
contains a minimal window of $H$. Let $V$ be a collection of minimal windows of
$H$ constructed by taking one minimal window from each window $w \in \nm{G;
s}$. It is obvious that the windows in $V$ do not overlap and that $\abs{V} =
\nm{G; s}$.  Proposition~\ref{prop:monotonic} implies that $\abs{\nm{H; s}}
\geq \abs{V}$.

\end{IEEEproof}

In~\cite{laxman:07:fast} the authors introduce a measure for the episodes to be
the maximal number of non-overlapping occurrences of the episode $s$. Since each
occurrence is either a minimal window or contains a minimal window,
Proposition~\ref{prop:monotonic} tells us that $\nm{G; s}$ is exactly this
measure.

\section{Experiments}
In this section we present our experiments with the quality
measure using synthetic and real-world text sequences.
\label{sec:experiments}
\subsection{Datasets}
We conducted our experiments with several synthetic and real-world sequences. 

The first synthetic sequence, \emph{gen-ind} consisted of $200000$ digits drawn
independently from the uniform model. The purpose of this dataset is to show
that our method finds very few significant episodes.  The second synthetic
sequence, \emph{gen-co} also consisted of $200000$ digits.  The sequence was
generated as follows. First we choose, by a fair coin flip, whether to generate
a digit from $0$ -- $4$ or $5$ -- $9$. In the former case the digit was selected
from a uniform model. In the latter case the probability of selecting the digit
$i$ was proportional to $0.5^x$, where $x$ is the distance between the current
location and the last location of the digit $i - 5$. Thus in this sequence, the
digits $i$ and $i - 5$ tend to be close to each other.

Our third dataset, \emph{moby}, was the novel Moby Dick by Herman
Melville.\footnote{The book was taken from
\url{http://www.gutenberg.org/etext/15}.}  Our fourth sequence, \emph{abstract}
consisted of 739 first NSF award abstracts from 1990.\footnote{The abstracts
were taken from \url{http://kdd.ics.uci.edu/databases/nsfabs/nsfawards.html}}
Our final dataset, \emph{address}, consisted of 
inaugural addresses of the presidents of the United States.\footnote{The
addresses were taken from~\url{http://www.bartleby.com/124/pres68}.} To
avoid the historic concept drift we entwined the speeches by first taking the
odd ones and then even ones.  The sequences were processed using the Porter
Stemmer and the stop words were removed.

\begin{table}[htb!]
\centering
\begin{tabular}{rllll}
\toprule
Sequence & Size & $\abs{\Sigma}$ & $N$ & $K$ \\
\midrule
\emph{gen-ind}  & 200000 & 10   & 4000 & 40 \\
\emph{gen-co}   & 200000 & 10   & 3500 & 35 \\
\emph{moby}     & 105719 & 10277 & 20   & 10 \\
\emph{abstract} & 67828  & 6718  & 22   & 10 \\
\emph{address}  & 62066  & 5295  & 20   & 10 \\
\bottomrule
\end{tabular}
\caption{Characteristics of the sequences and the threshold values used for
mining candidate episodes. The second column is the number of symbols in the
sequence. The third column is the threshold for the number of minimal windows
and the fourth column is the largest minimal window considered.}
\label{tab:basic}
\end{table}

\subsection{Experimental Setup}
Our experimental setup mimics the framework setup in~\cite{webb:07:discovering}
in which the data is divided into two parts, the first part is used for
discovering the patterns and the second part for testing whether the discovered
patterns were significant. We divided each sequence into two parts of
equivalent lengths. We used the first sequence for discovering the candidate
episodes and training the independence model. Then the discovered episodes
were tested against the model using the second sequence.

As candidate episodes we considered only those episodes whose number of
non-overlapping windows exceeded some threshold $N$.  When computing the
independence model and discovering minimal windows from the test data
we only considered the minimal windows of at most $K$.  We used $K
= 10$ for the text sequences and $K = 35, 40$ for the synthetic sequences.
These thresholds are given in Table~\ref{tab:basic}.

Let $\efam{G}$ be the set of candidates. Since we compute the samples from the
test sequence, it is not guaranteed that an episode $G \in \efam{G}$ will have
enough minimal windows. Hence we discard any episode whose number of minimal
windows in the test sequence does not exceed $N$. We also remove any episodes
having the variance $0$ since for these episodes the minimal window will always
be of the same known size. This set includes all singletons. Let us denote this
set of episodes by $\efam{H}$. The sizes of these families along with the sizes
of the machines $\simple{\mach{\efam{H}}}$ and $\co{\mach{\efam{H}}}$ are given
in Table~\ref{tab:setup}.

For each episode $H \in \efam{H}$ we computed the $Z$-statistic given in
Eq.~\ref{eq:ztest}. This value is asymptotically distributed as a standard
normal distribution. We considered two different $P$-values. First, we computed
a one-sided $P$-value to we examine whether $Z$ is abnormally small, thus our
test will return small $P$-values if the minimal windows are significantly
smaller than expected. Secondly, we computed a two sided $P$-value to test
whether the average of lengths of minimal windows is significantly smaller or
larger. The correlation between the minimal windows (see
Section~\ref{sec:test}) was computed by simulating a sequence with $10^6$
symbols. The computation of $P$-values lasted about $5$ minutes for the
generated sequences and less than a minute for text sequences.  The most
expensive step was the computation of the correlation terms explained in
Section~\ref{sec:test}.

\begin{table}[htb!]
\centering
\begin{tabular}{rllll}
\toprule
Sequence & $\abs{\efam{G}}$ & $\abs{\efam{H}}$ & $\abs{\simple{\mach{\efam{G}}}}$ & $\abs{\co{\mach{\efam{G}}}}$ \\
\midrule
\emph{gen-ind}  & 4882  & 4872 & 4889  & 28046 \\
\emph{gen-co}   & 3993  & 3982 & 4221  & 24035 \\
\emph{moby}     & 724   & 137  & 726   & 2382 \\
\emph{abstract} & 14569 & 116  & 14985 & 106901 \\
\emph{address}  & 482   & 78   & 483   & 1551 \\
\bottomrule
\end{tabular}
\caption{Sizes of data structures in experiments. The first column is the
number of candidate episodes, the second column is the number of episodes
actually tested. The third column is the number of states in
$\simple{\mach{\efam{G}}}$ and the fourth column is the number of states in
$\co{\mach{\efam{G}}}$.}
\label{tab:setup}
\end{table}

\subsection{Significant Episodes}
In this section we will focus on the episodes discovered by our approach.

From each candidate set we computed the significant episodes based on their
$P$-values. As a significance level we used $0.05$. We compared raw $P$-values
and also adjusted $P$-values. The adjustment was done using the Benjamini
Hochberg Procedure in order to control the FDR family-wise
error~\cite{benjamini:01:control}. The results are given in
Table~\ref{tab:results}.

\begin{table}[htb!]
\centering
\begin{tabular}{rlllll}
\toprule
& \multicolumn{2}{c}{Raw} && \multicolumn{2}{c}{Adjusted} \\
\cmidrule{2-3}
\cmidrule{5-6}
Sequence & one-s. & two-s. && one-s. & two-s. \\
\midrule
\emph{gen-ind}  & 446 & 355 && 0 & 0 \\
\emph{gen-co}   & 237 & 101 && 242 & 90  \\
\emph{moby}     & 23 & 20 && 12 & 9 \\
\emph{abstract} & 41 & 42 && 15 & 15 \\
\emph{address}  & 20 & 19 && 3 & 3 \\
\bottomrule
\end{tabular}
\caption{Significant episodes according to their raw and adjusted $P$-values.
Significance level is $0.05$. The $P$-values were adjusted with the Benjamini
Hochberg Procedure method in order to control the FDR error.  }
\label{tab:results}
\end{table}

Let us first consider \emph{gen-ind}. Since this sequence correspond to the
independence model there should be no significant episodes. However, since we
are accepting $5\%$ of false significant episodes we should expect about $240$
significant episodes.  The number of significant episodes discovered is about
$10\%$. The higher number for these tests can be explained by the fact that the
model we are using is actually trained from the training data and hence contain
some error. Should we use the exact model, then the number of significant
episodes will drop to $5\%$. After adjusting the raw $P$-values, no significant
episode remained. Hence, our method did not find any significant episode from
\emph{gen-ind}, as expected.

Next we will consider the sequence \emph{gen-co}. Here we expect to find
significant patterns, since the sequence does not obey the independence model.
We see from Table~\ref{tab:results} that this is the case. Even after the
adjustment there is a considerate amount of significant episodes. By studying
the results we found out that the significant episodes had either the basic
form of $i \to i + 5$ or a combination of these. This is an expected result
since the sequence had $i$ and $i + 5$ abnormally close to each other. An
important observation here is that the algorithm also discovers complex
episodes to be important. Namely, the $P$-value, our quality measure  is fair
for simple and more complex episodes.

Our next sequence was \emph{moby}. Since the alphabet in this sequence is quite
large, the number of candidate episodes is rather small and a lot of these
candidate episodes are in fact singletons --- in the end $137$ episodes were
given a $P$-value. Out of these episodes about $15\%$ were significant based on
raw $P$-value and about $10\%$ when $P$-values were adjusted.  Some examples
among the most significant episodes based on one-sided test were $(white \to
whale)$, $(sperm \to whale)$, $(old \to man)$, along with their parallel
versions.  Such episodes imply that these words occur abnormally close to each
other. On the other hand, candidate episodes such as $(time, whale)$ or $(ship,
man)$ were not considered significant. This means that even though these
combinations occur often, the episode can be explained by the fact that their
individual words are common. Similarly, some of the significant episodes
discovered in \emph{abstract} were $(research \to project)$ and
$(undergraduate, student)$. Episodes discovered from \emph{address} were
$(united \to states)$, $(united, states)$ and $(fellow, citizen)$.

\section{Related Work}
\label{sec:related}
Our approach resembles the approach taken
in~\cite{gwadera:05:reliable,gwadera:05:markov} in which the authors considered
episode to be significant if the episode occurs too often or not often enough
in a fixed window. As a background model the authors used independence model
in~\cite{gwadera:05:reliable} and markov-chain model
in~\cite{gwadera:05:markov}. The main difference between our approach and
theirs is that we are studying the behavior of the minimal windows. As we have
discussed in the introduction we believe that using the statistics based on
minimal windows has an advantage over the fixed window approach.

In~\cite{casas-garriga:03:discovering}, the author proposed a criterion for
episodes by requiring that the consecutive symbols in a sequence should only
within a specified bound. While this approach attacks the problem of fixed
windows, it is still a frequency-based measure. This measure, however, is not
antimonotonic as it is pointed out in~\cite{meger:04:constraint-based}.  It
would be useful to see whether we can compute an expected value of this measure
so that we can compute a $P$-value based on some background model.

In a related work~\cite{cule:09:new} the authors considered parallel episodes
significant if the smallest window containing each occurrence of a symbol of an
episode had a small value. Their approach differ from ours since the smallest
window containing a fixed occurrence of a symbol is not necessarily the minimal
window. Also, they consider only parallel episodes whereas we consider more
general DAG episodes. An interesting approach has been also taken
in~\cite{calders:07:mining} where the authors define a windowless frequency
measure of an itemset within a stream $s$ to be the frequency starting from a
certain point. This point is selected so that the frequency is maximal.
However, this method is defined for itemsets and it would be fruitful to see
whether this idea can be extended into episodes.

Finite state machines have been used
in~\cite{troncek:01:episode,hirao:01:practical} for discovering episodes.
However, their goal is different than ours since the actual machine is built
upon a sequence and not the episode set and it is used for discovering episodes
and not computing the coverage.

\section{Discussion and Conclusions}
\label{sec:conclusions}
In this paper we proposed a new quality measure for the episodes. Our approach
tackles simultaneously problems with fixed windows but also allows us to
incorporate background knowledge. The measure itself is a deviation of the
average length of the minimal windows when compared to the expected length
according to the independence model.

Our main technical contribution is the technique for computing the distribution
of lengths of minimal window. In order to do that we create an elaborate finite
state machine and compute probabilities iteratively starting from simple
episodes and moving toward complex ones. Once the distribution is computed we
are able to perform a statistical test on the discovered minimal windows from the
test sequence. Our experiments with the text data suggest that this measure
finds significant episodes while ignoring uninteresting ones.

The proposed method requires a parameter $K$, a limit to the size of a minimal
window. In this paper we simply have assumed that this parameter is
domain-specific and is provided by the user. Setting this parameter high may
allow us to discover more interesting patterns. However, when using large
values for $K$,  computing the model may become computationally infeasible as
we are forced to use exact rational numbers in order to guarantee numerical
stability. This computational problems may be solved by simulating the
independence model instead of computing the exact probabilities. In such case,
more analysis is needed to determine a proper number of steps in this
simulation.

As a future work we also consider more elaborate models such as Markov Chains.
This has been done in~\cite{gwadera:05:markov} for windows of fixed size and
our goal is to extend this approach for minimal windows.

Our experiments revealed an interesting behavior within certain sets. Certain
information tend to repeat in several forms of episodes. For example, we found
that both $(white, whale)$ and $(white \to whale)$ were significant. This
suggests that there is a need for pattern reduction techniques. Such techniques
are well studied in the setting of itemsets but are not that well developed for
episodes.

\bibliographystyle{IEEEtran}
\bibliography{bibliography}

\end{document}